\documentclass[aps,pra,showpacs,twocolumn,superscriptaddress,amsmath,amssymb]{revtex4}
\usepackage{graphicx, graphics}
\usepackage{amsmath}
\usepackage{color}
\begin{document}

\title{Supersolid with nontrivial topological spin textures in spin-orbit-coupled Bose gases}
\author{Wei Han}
\affiliation{Beijing National Laboratory for Condensed Matter
Physics, Institute of Physics, Chinese Academy of Sciences, Beijing
100190, China}
\author{Gediminas Juzeli\={u}nas}
\affiliation{Institute of Theoretical Physics and Astronomy, Vilnius
University, A. Go\v{s}tauto 12, Vilnius 01108, Lithuania}
\author{Wei Zhang\footnote{wzhangl@ruc.edu.cn}}
\affiliation{Department of Physics, Renmin University of China,
Beijing 100872, China}
\author{Wu-Ming Liu\footnote{wliu@iphy.ac.cn}}
\affiliation{Beijing National Laboratory for Condensed Matter
Physics, Institute of Physics, Chinese Academy of Sciences, Beijing
100190, China}

\begin{abstract}
Supersolid is a long-sought exotic phase of matter, which is
characterized by the co-existence of a diagonal long-range order of
solid and an off-diagonal long-range order of superfluid. Possible
candidates to realize such a phase have been previously considered,
including hard-core bosons with long-range interaction and soft-core
bosons. Here we demonstrate that an ultracold atomic condensate of
hard-core bosons with contact interaction can establish a supersolid
phase when simultaneously subjected to spin-orbit coupling and a
spin-dependent periodic potential. This supersolid phase is
accompanied by topologically nontrivial spin textures, and is
signaled by the separation of momentum distribution peaks, which can
be detected via time-of-flight measurements. We also discuss
possibilities to produce and observe the supersolid phase for
realistic experimental situations.
\end{abstract}

\pacs{03.75.Lm, 03.75.Mn, 67.85.Hj, 67.80.K-} \maketitle

\section{Introduction}
The search for supersolid phase has a long history since
1969~\cite{Andreev,Chester,Leggett,Meisel,Balibar,Boninsegni}, and
has been recently intensified during the debate of its possible
observation in $^{4}$He~\cite{Balibar2,Kim,Ye,Day,Hunt,Kim2,Chan}.
From the theoretical aspect, it has been suggested that supersolid
can exist in condensates of soft-core bosons
\cite{Henkel,Cinti,Henkel2} and hard-core bosons with long-range
interactions \cite{Troyer,Danshita,Tieleman}. However, the
realization of supersolid in hard-core bosons with short-range
interactions is usually considered unlikely \cite{Boninsegni2}.

Thanks to the high controllability, ultracold atomic gases provide
us an excellent platform to emulate various quantum phenomena
originally considered in the context of condensed matter
physics~\cite{Lewenstein,Bloch}. Recent experimental realizations of
artificial spin-orbit (SO) coupling
\cite{Spielman,Shuai-Chen,Shuai-Chen2,Jing-Zhang,Zwierlein,Engels,Jing-Zhang2}
introduce another degree of freedom for the manipulation of atomic
gases, and give opportunities for the search of novel quantum states
\cite{Dalibard,Galitski,Wu,Niu,Galitski2,Wu2,Wuming-Liu,
Trivedi,Wei-Zhang}. Theoretical investigations reveal that the
interplay among the SO coupling, interatomic interaction and
external potential can lead to diverse phase diagrams for Bose
gases, containing the plane wave, density stripe, composite soliton,
vortex lattice, as well as quantum quasicrystal
\cite{Hui-Zhai,Tin-Lun-Ho,Malomed,You,Hu,Santos,
Gou,Su-Yi,Gopalakrishnan}. The SO-coupled ultracold atomic gas is
also opening new perspectives in the supersolid phenomena
\cite{Ye2,Pitaevskii}.

In this manuscript we investigate a hard-core Bose gas interacting
via a contact (zero-range) potential. The atoms experience a
spin-dependent periodic potential \cite{Deutsch,Mandel} and are
subjected to two-dimensional (2D) SO coupling of the
Rashba-Dresselhaus type
$\mathcal{V}_{\text{so}}=-i\hbar(\kappa_{x}\sigma_{x}\partial_{x}
+\kappa_{y}\sigma_{y}\partial_{y})$~\cite{Goldman}. Here,
$\sigma_{x,y}$ are the Pauli matrices and $\kappa_{x,y}$ represent
the corresponding SO-coupling strengths. We demonstrate that a
supersolid phase characterized by the coexistence of periodic
density modulation and superfluidity can be stabilized by strong SO
coupling. Comparing to a continuous system affected by SO coupling
discussed in Ref.~\cite{Pitaevskii}, the supersolid phase in the
present system involving a spin-dependent periodic potential is
accompanied by the spontaneous generation of a lattice composed of
meron pairs and antimeron pairs, hence featuring a topologically
nontrivial spin configuration. With decreasing the SO-coupling
strength, this supersolid phase gives way to a state consisting
of alternating spin domains separated by chiral Bloch walls.
Depending on the sign of SO coupling, the chirality of the Bloch
walls can be either right-handed or left-handed. We also discuss the
influence of asymmetric interatomic interaction and SO coupling
anisotropy $(\kappa_x \neq \kappa_y)$ on the properties of the
supersolid phase.
\begin{figure*}[tbp]
\includegraphics[width=13.5cm]{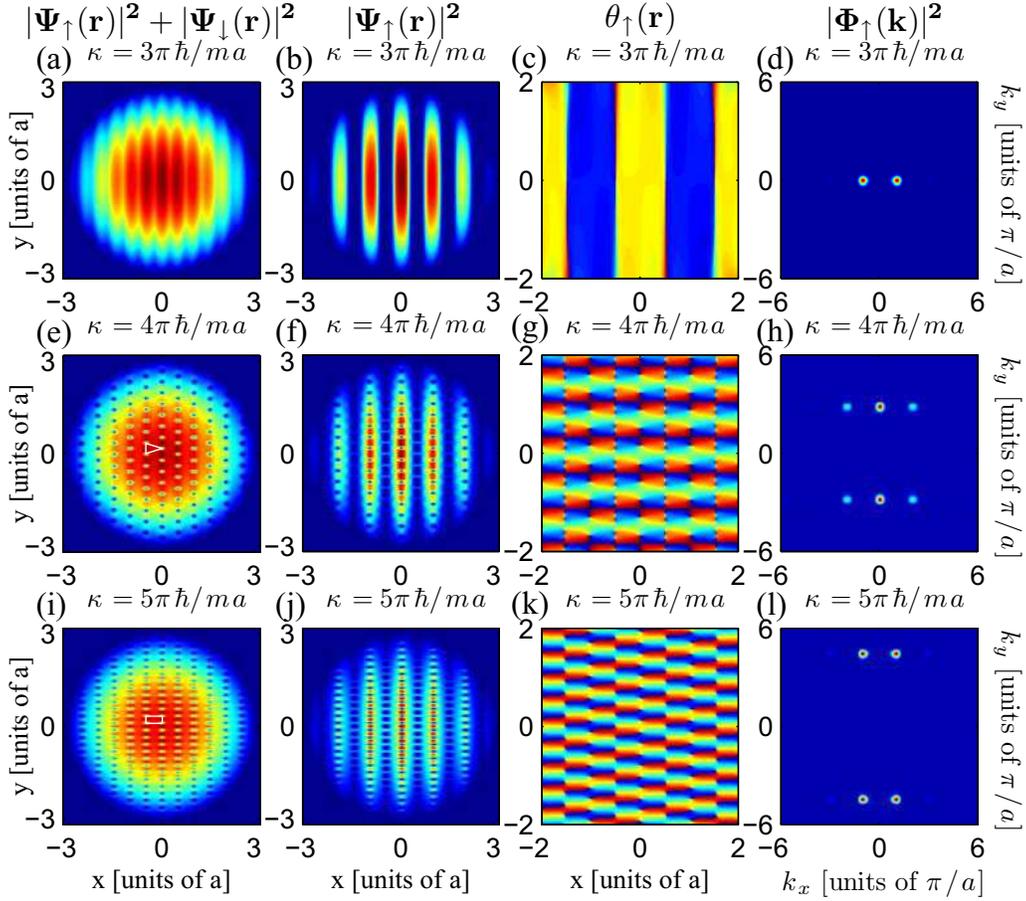}
\caption{(Color online) Spin-orbit-coupling induced transition from
superfluid to supersolid. Various ground-state properties of Rashba
SO-coupled Bose gases loaded in a 1D spin-dependent periodic
potential are shown, with the SO-coupling strength
$\kappa=3\pi\hbar/ma$ (top), $4\pi\hbar/ma$ (middle), and
$5\pi\hbar/ma$ (bottom). With increasing SO-coupling strength, the
system undergoes a phase transition from the superfluid phase (top)
to a supersolid phase (middle and bottom) characterized by density
modulation along the $y$ direction (first and second columns). The
supersolid phase features a triangular lattice (middle) and a
rectangular lattice (bottom) at intermediate and high SO-coupling
strengths, respectively. The supersolid is accompanied by a
spontaneous generation of vortex and antivortex chains in the
spin-up and spin-down domains, respectively, as can be seen from the
condensate phase modulation (third column). The occurrence of the
supersolid phase is signaled by the separation of the momentum
distribution peaks (fourth column), and can be readily observed via
time-of-flight measurements. Other parameters used in these plots
are $V_{0}=20\pi^2\hbar^2/ma^2$ and $\tilde{g}=6000$.} \label{fig1}
\end{figure*}

\section{Spin-Orbit-Coupling Induced Supersolid}
We consider SO-coupled two-component Bose-Einstein condensates in a
spin-dependent periodic potential. The SO coupling is of the
Rashba-Dresselhaus type, which can be realized by using a periodic
pulsed magnetic field~\cite{Anderson,Xu}. The spin-dependent
periodic potential is usually produced by means of the
counterpropagating cross-polarized laser
beams~\cite{Deutsch,Mandel}. Simultaneous creation of SO
coupling and spin-dependent periodic potential is discussed in
Appendix \ref{App:Creat}. The Hamiltonian reads in the
Gross-Pitaevskii mean-field approximation as
\begin{eqnarray}
\mathcal{H} &=&\int d\mathbf{r\Psi }^{\dag}\left(-\frac{\hbar ^{2}\boldsymbol{%
\nabla }^{2}}{2m}+\mathcal{V}_{\text{so}}\right)\mathbf{\Psi }+\int d\mathbf{r}%
\sum\limits_{\alpha =\uparrow ,\downarrow }V_{\alpha }\Psi
_{\alpha }^{\ast }\Psi _{\alpha }  \notag \\
&&+\frac{1}{2}\int d\mathbf{r}\sum\limits_{\alpha ,\beta =\uparrow
,\downarrow }g_{\alpha \beta }\Psi _{\alpha }^{\ast }\left( \mathbf{r}%
\right) \Psi _{\beta }^{\ast }\left( \mathbf{r}\right) \Psi _{\beta
}\left( \mathbf{r}\right) \Psi _{\alpha }\left( \mathbf{r}\right)
,\label{Model Hamiltonian}
\end{eqnarray}
where the complex-valued order parameter $\mathbf{\Psi}
=[\Psi_{\uparrow}(\mathbf{r}),\Psi_{\downarrow}(\mathbf{r})]^\top$
is normalized to the total particle number $N$ as $\int d\mathbf{r}
\mathbf{\Psi}^{\dag}\mathbf{\Psi}=~N$. The strength of the atom-atom
interaction $g_{\alpha\beta}=4\pi\hbar^{2}a_{\alpha\beta}/m$ is
characterized by the $s$-wave scattering length $a_{\alpha\beta}$.
The SO-coupling term can be written as
$\mathcal{V}_{\text{so}}=-i\hbar(\kappa_{x}\sigma_{x}\partial_{x}+\kappa_{y}\sigma_{y}\partial_{y})$,
where $\sigma_{x,y}$ are the Pauli matrices and $\kappa_{x,y}$
denote the SO-coupling strengths. In the isotropic case when
$\kappa_x = \kappa_y$, the SO coupling belongs to the Rashba type.
The spin-up and spin-down atoms are subjected to the spin-dependent
periodic potentials $V_{\uparrow}=V_{0}\sin ^{2}(\pi x/a)$ and
$V_{\downarrow}=V_{0}\cos ^{2}(\pi x/a)$, respectively.

The many-body ground state can be obtained by numerically minimizing
the Hamiltonian functional given by Eq.~(\ref{Model Hamiltonian}),
as outlined in Appendix \ref{App:Calculat}. In our calculation, we
additionally introduce a weak harmonic trap $V_{\text{H}}=m[\omega
_{\perp }^{2}(x^{2}+y^{2})+\omega_{z}^2z^{2}]/2$ with $\omega
_{\perp}=\pi^{2}\hbar/ma^{2}$ to simulate realistic configurations
of cold atom experiments. When $\lambda=\omega_{z}/\omega_{\perp}
\gg~1$, the condensates can be regarded as quasi-2D, and the
effective interaction parameter in a 2D dimensionless form is
$\tilde{g}_{\alpha\beta}=2\sqrt{2\pi}Na_{\alpha\beta}/a_{hz}$, where
$a_{hz}=\sqrt{\hbar/m\omega_{z}}$. Considering that the differences
in $a_{\uparrow\uparrow}$, $a_{\downarrow\downarrow}$ and
$a_{\uparrow\downarrow}$ are within $1\%$ in typical experiments
involving the magnetic sublevels of alkali atoms, first we focus on
the case of SU(2) symmetric interactions with
$\tilde{g}=\tilde{g}_{\uparrow\uparrow}=\tilde{g}_{\downarrow\downarrow}=\tilde{g}_{\uparrow\downarrow}$.

\begin{figure}[t]
\includegraphics[width=8cm]{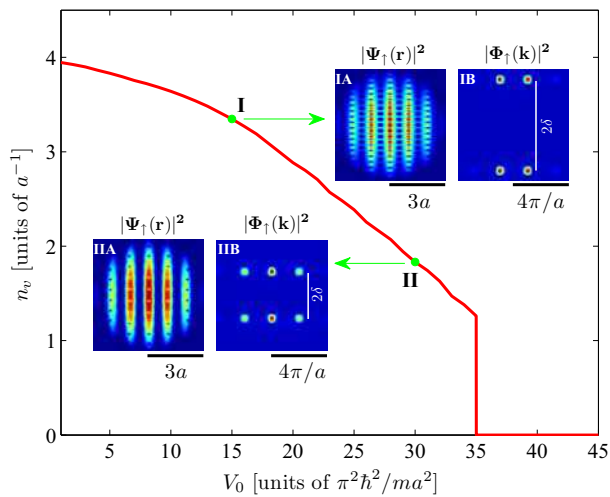}
\caption{(Color online) Line density of vortices as a function of
periodic potential depth. With increasing the depth $V_{0}$ of the
periodic potential, the line density $n_{v}$ of the vortices
decreases gradually, and drops to zero suddenly at the transition
point $V_{0}=35\pi^2\hbar^2/ma^2$ between the supersolid and
superfluid phases. The insets depict the change of vortex density
and atomic momentum distributions. The line density $n_{v}$ of the
vortices is proportional to the separation $\delta$ of the momentum
distribution peaks, and can be expressed as $n_{v}=\delta/\pi$. The
Rashba SO-coupling strength is fixed at $\kappa=4\pi\hbar/ma$, and
the dimensionless interaction parameter is taken as
$\tilde{g}=6000$.} \label{fig2}
\end{figure}

For a fixed value of atom-atom interaction, we observe a transition
from the superfluid phase to a supersolid phase with increasing
Rashba SO-coupling strength $\kappa$, as one can see in
Fig.~\ref{fig1}. Specifically, when the SO coupling is weak, the
ground state of the system consists of alternating spin domains,
where stripes filled with spin-up and spin-down atoms are
segregated~[see Figs.~\ref{fig1}(a) and \ref{fig1}(b)]. While the translational
symmetry along the $x$ direction is explicitly broken by the
spin-dependent periodic potential, the system preserves its
translational symmetry along the $y$ direction. If the strength of
the SO coupling is increased beyond a critical value, the
translational symmetry along the $y$ direction is spontaneously
broken. As a result, a new phase with periodic density modulation
along the $y$ direction is stabilized~[see Figs.~\ref{fig1}(e)-\ref{fig1}(f) and
\ref{fig1}(i)-\ref{fig1}(j)], and hence can be considered as a supersolid state. The
emergence of such a density modulation can be understood as a stripe
phase along the $y$ direction induced by SO coupling. However, this
supersolid phase also exhibits exotic spin textures, which will be
discussed below. The momentum distribution of the supersolid phase
features a qualitative difference from the superfluid phase. In the
superfluid phase, the atoms are condensed at a set of discrete
points on the edge of Brillouin zones with finite momenta $(k_{x}
\in K,k_{y}=0)$, where $K=\{\pm \pi/a,\pm 3\pi/a,\pm
5\pi/a,...\}$~[see Fig.~\ref{fig1}(d)]. In the supersolid phase, the
momentum distribution peaks are separated from $k_{y}=0$ to
$k_{y}=\pm \delta $. The separation distance $\delta \in (0,
m\kappa/\hbar)$ depends on the SO-coupling strength [see
Figs.~\ref{fig1}(h) and \ref{fig1}(l)] and the periodic potential
depth [see Fig. \ref{fig2}]. This qualitative difference can be
detected using conventional time-of-flight imaging technique.

\begin{figure}[tbp]
\includegraphics[width=6.3cm]{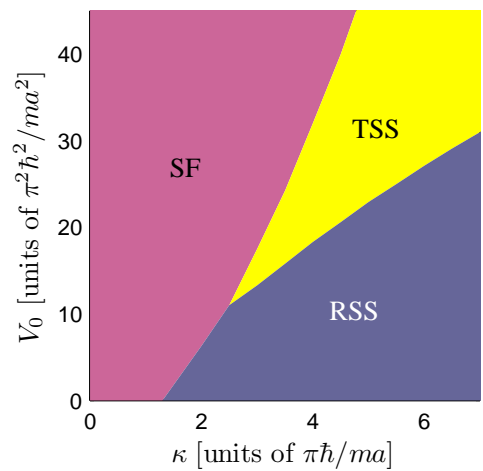}
\caption{(Color online) The ground-state phase diagram spanned by
the Rashba SO-coupling strength $\protect\kappa$ and the periodic
potential depth $V_{0}$. Three phases can be identified on this
phase diagram, including the superfluid (SF) phase, the triangular
supersolid (TSS) phase and the rectangular supersolid (RSS) phase.
The dimensionless interaction parameter is taken as
$\tilde{g}=6000$.} \label{fig3}
\end{figure}

In addition to the density modulation along the $y$ direction, the
supersolid phase is also characterized by a vortex lattice structure
consisting of vortex and antivortex chains in the spin-up and
spin-down domains, respectively [see Figs.~\ref{fig1}(e)-\ref{fig1}(g) and \ref{fig1}(i)-\ref{fig1}(k)].
Depending on the competition between the SO-coupling strength and
periodic potential depth, two different arrangements of vortices can
be stabilized. In one case, the vortices of the neighboring chains
are staggered, forming a triangular lattice~[Fig.~\ref{fig1}(e)]. In
the other case, the vortices of the neighboring chains are parallel,
forming a rectangular lattice~[Fig.~\ref{fig1}(i)]. As shown in
Figs. \ref{fig1}(h) and \ref{fig1}(l), these two different vortex lattices correspond
to qualitatively different momentum distributions, and hence can be
distinguished by experiments. In Fig. \ref{fig3}, we present the
ground-state phase diagram spanned by the SO-coupling strength
$\kappa$ and the periodic potential depth $V_{0}$, with the
effective interaction parameter being $\tilde{g}=6000$.
\begin{figure*}[tbp]
\includegraphics[width=13.5cm]{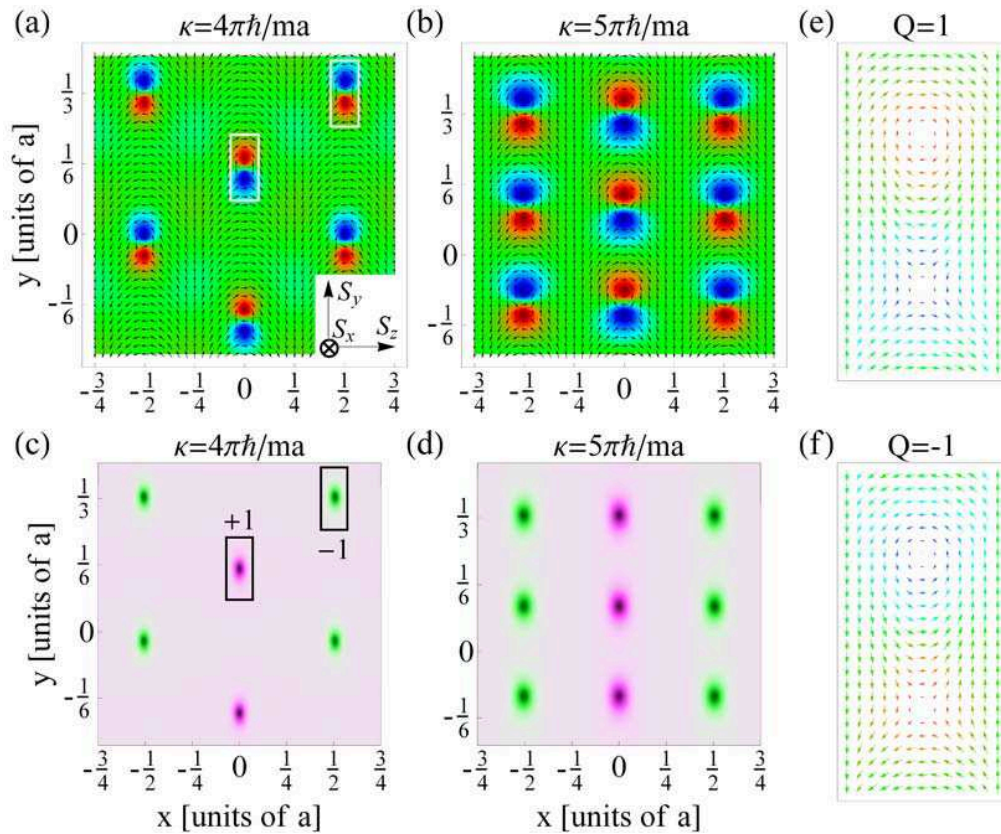}
\caption{(Color online) Topological spin textures. (a), (b) Spin
configurations of the triangular and rectangular supersolid phases
showing the spontaneous emergence of meron pairs and antimeron
pairs. In these plots, the arrows represent the directions of spin
vector ($S_{z}$, $S_{y}$) and the colors ranging from blue to red
describe the values of $S_{x}$ from $-1$ to $1$. (c), (d) Density
distributions $q(\mathbf{r})$ of topological charge for the spin
textures shown in (a) and (b). The pink and green bubbles indicate
that the meron pairs and antimeron pairs carry positive and negative
topological charges, respectively. (e), (f) Schematic spin
configurations of a meron pair (e) and an antimeron pair (f). Both
the meron pairs and antimeron pairs have the same
``circular-hyperbolic" structure, but with opposite spin
orientations. Parameters used in these plots are identical to those
used in Figs. \ref{fig1}(e)-\ref{fig1}(h) and~\ref{fig1}(i)-\ref{fig1}(l).} \label{fig4}
\end{figure*}

We stress that a vortex lattice is not directly associated with the
supersolid phase, as it is absent in the supersolid droplet crystals
\cite{Cinti}. In the present system, the generation of vortices is a
direct consequence of the interplay between the SO coupling,
spin-dependent periodic potential, and interatomic interactions.
This is very different from the usual manner of creating supersolid
vortices by rotation \cite{Henkel2} or artificial magnetic fields
\cite{Tieleman}.

The alternating arrangement of vortex and antivortex chains can be
viewed as alternating plane waves propagating on opposite directions
along the $y$ axis, as shown in Figs.~\ref{fig1}(g) and \ref{fig1}(k). According to
the
Onsager-Feynman quantization condition \cite{Pethick} $\oint\nolimits_{\mathcal{C}}\mathbf{v}_{\text{s}%
}\cdot d\mathbf{l}=2\pi \hbar N_{\text{v}}/m$, we can express the
line density of the vortices as $n_{\text{v}}=k_{y}/\pi$, where
$k_{y}=\delta$ is the wave number of the plane waves. Numerical
simulations show that for a given SO-coupling strength $\kappa$ the
line density of vortices decreases from $m\kappa/\pi \hbar$ to $0$
with increasing the periodic potential depth $V_{0}$, as one can see
in Fig. \ref{fig2}.

\section{Topological spin textures}
The two-component Bose gas can be considered as a magnetic system.
Thus one might naturally think that the supersolid transition would
be associated with some magnetic ordering. We next demonstrate that
the supersolid indeed features topologically nontrivial spin
textures. To see this, we define a spin density vector
$\mathbf{S}=\mathbf{\Psi}
^{\dag}\boldsymbol{\sigma}\mathbf{\Psi}/|\mathbf{\Psi}|^{2}$ in the
pseudospin representation, where $\boldsymbol{\sigma}$ is the Pauli
matrix vector. Vectorial plots of $\mathbf{S}$ (under a pseudospin
rotation $\sigma _{x}\rightarrow \sigma _{z}$ and $\sigma
_{z}\rightarrow -\sigma _{x}$) are shown in Figs.~\ref{fig4}(a) and \ref{fig4}(b) for
the triangular and rectangular lattices, respectively, where the
parameters are the same as in Figs.~\ref{fig1}(e) and \ref{fig1}(i). In both
Figs.~\ref{fig4}(a) and \ref{fig4}(b), the spin texture represents a spontaneous
magnetic ordering in the form of crystals of meron pairs and
antimeron pairs~\cite{Volovik}. The meron pairs reside in the
spin-up domains, while the antimeron pairs reside in the spin-down
ones. We note that a meron is a topological configuration in which
the spin points up or down at the meron core and rotates away from
the core. Both a meron pair and an antimeron pair have a
``circular-hyperbolic" structure, and the only difference is that
they have exactly opposite spin orientations [see Figs.~\ref{fig4}(e) and
\ref{fig4}(f)].

The topological nature of the spin textures can be characterized by
the topological charge $Q$ (Chern number), which is defined as a
spatial integral of the topological charge density $q\left(
\mathbf{r}\right) =(1/8\pi) \epsilon ^{ij}\mathbf{S}\cdot
\partial _{i}\mathbf{S} \times \partial _{j}\mathbf{S}$. In Figs.
\ref{fig4}(c) and \ref{fig4}(d), we present the topological charge density
distribution for the triangular and rectangular lattices,
respectively. Notice that both meron pairs and antimeron pairs are
topologically nontrivial. A meron pair carries a topological charge
$1$, while an antimeron pair carries a topological charge $-1$. As a
comparison, the topological charge density is zero everywhere in the
topologically trivial superfluid phase.

Topological spin texture lattices, such as meron-pair and skyrmion
lattices, are usually stabilized by bulk rotation
\cite{Kasamatsu,Schweikhard,Gou2}. Recently, it has been also
suggested that skyrmion lattices can be realized by the combined
effects of SO coupling and harmonic trap \cite{Hu}, provided that
the trapping potential energy $\hbar\omega_{\perp}$ is higher than
the characteristic interaction energy $\tilde{g}\hbar\omega_{\perp}$
(i.e., $\tilde{g}<1$)~\cite{Santos}. Our results demonstrate that
meron-pair lattices can also be stabilized by the combined effects
of SO coupling and spin-dependent periodic potential, within a large
regime of interatomic interaction strength. For example, the
meron-pair lattices in Figs.~\ref{fig4}(a) and \ref{fig4}(b) are obtained at the large
effective interaction strength with $\tilde{g}=6000 \gg 1$. This
interaction strength can be naturally realized in experiments under
realistic conditions without resort to the Feshbach resonance. This
observation hence provides a way to create and manipulate
topological spin textures in SO-coupled systems.
\begin{figure}[tbp]
\includegraphics[width=8.0cm]{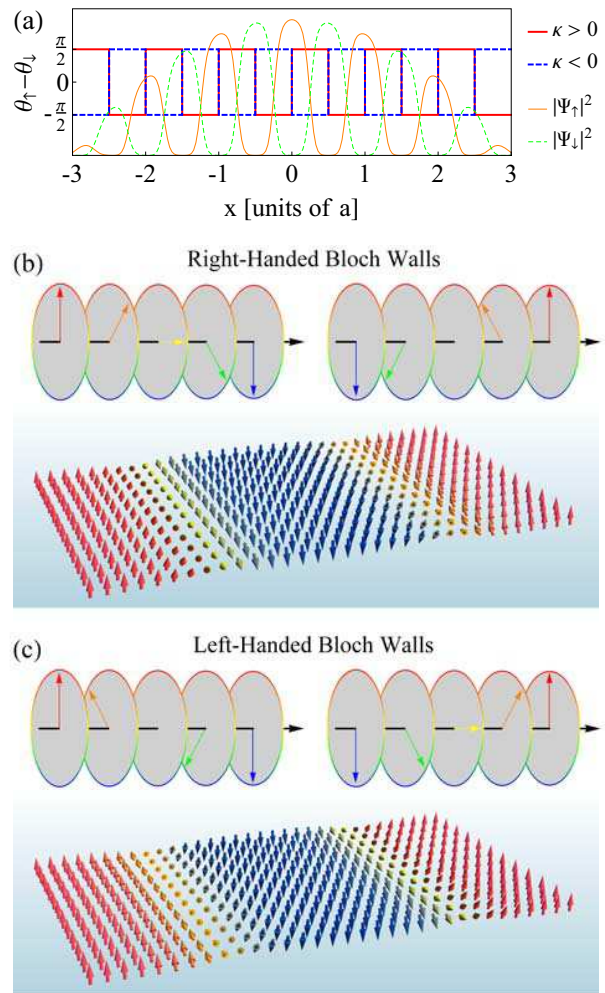}
\caption{(Color online) Domain wall chirality. (a) Section view of
the relative phase $\theta_{\uparrow}-\theta_{\downarrow}$ along the
$x$ axis. The presence of Rashba SO coupling locks the relative
phase at $\pm\pi/2$, and the periodic change of the densities along
the $x$ direction induces a periodic modulation of the relative
phase between $\pi/2$ and $-\pi/2$. By tuning the sign of the Rashba
SO coupling from positive ($\kappa>0$) to negative ($\kappa<0$), the
relative phase will be changed from $\pm\pi/2$ (red thick line) to
$\mp\pi/2$ (blue thick dashed line). (b) An illustration of
right-handed chiral Bloch walls stabilized by positive Rashba SO
coupling with $\kappa>0$. (c) An illustration of left-handed chiral
Bloch walls stabilized by negative Rashba SO coupling with
$\kappa<0$. By crossing a right-handed (left-handed) chiral Bloch
wall, the spin vector flips like a right-rotating (left-rotating)
spiral.} \label{fig5}
\end{figure}

\section{Chiral domain walls}
After discussing the novel properties of the supersolid state in the
previous sections, here we investigate the superfluid phase appearing
at weak SO coupling. In this phase, the translational symmetry along
the $y$ axis (orthogonal to the direction of the 1D periodic
potential) is preserved, such that the system does not support
density modulation along this axis. However, the presence of SO
coupling breaks the spin-rotational symmetry in the $S_{x}$-$S_{y}$
plane, and leads to spontaneous chiral domain walls.

In order to give a clear description of this phenomenon, we first
consider the effect of SO coupling on the relative phase
$\theta_{\uparrow}-\theta_{\downarrow}$ of the two component
condensates, where $\theta _{\uparrow}$ and $\theta _{\downarrow}$
represent the phases of the spin-up and spin-down wave functions,
respectively. In the absence of SO coupling, the Hamiltonian of
Eq.~(\ref{Model Hamiltonian}) does not depend on the relative phase
between the two spin components. In the presence of SO coupling, for
the superfluid phase illustrated in
Figs.~\ref{fig1}(a)-\ref{fig1}(d), the phase of the components does
not alter except periodic jumps in the $x$ direction [see
Fig.~\ref{fig1}(c)], thus we have
$\nabla\theta_{\uparrow}=\nabla\theta_{\downarrow}=0$. Due to the
translational symmetry, the gradient of the density along the $y$
direction can be approximately considered as
$\partial_{y}|\Psi_{\uparrow}|^{2}\simeq\partial_{y}|\Psi_{\downarrow}|^{2}\simeq0$,
so the SO-coupling term in Eq.~(\ref{Model Hamiltonian}) can be
represented as
\begin{eqnarray}
\int\mathbf{\Psi}^{\dag}\mathcal{V}_{\text{so}}\mathbf{\Psi}
d\mathbf{r}=2\kappa\int
|\Psi_{\downarrow}|\partial_{x}|\Psi_{\uparrow}|\sin(\theta_{\uparrow}-\theta_{\downarrow})
d\mathbf{r}. \label{SO coupling term}
\end{eqnarray}
One can easily see that in the presence of SO coupling, the
Hamiltonian depends on the relative phase
$\theta_{\uparrow}-\theta_{\downarrow}$. By minimizing the energy
functional, the relative phase of the ground-state wave functions
has to be locked at $\pm\pi/2$, where the sign $\pm$ is determined
by the sign of $\partial_{x}|\Psi_{\uparrow}|$. As a result, the
periodic density modulation along the $x$ direction leads to a
relative phase alternating between $\pi/2$ and $-\pi/2$ [see
Fig.~\ref{fig5}(a)].
\begin{figure*}[tbp]
\includegraphics[width=13.5cm]{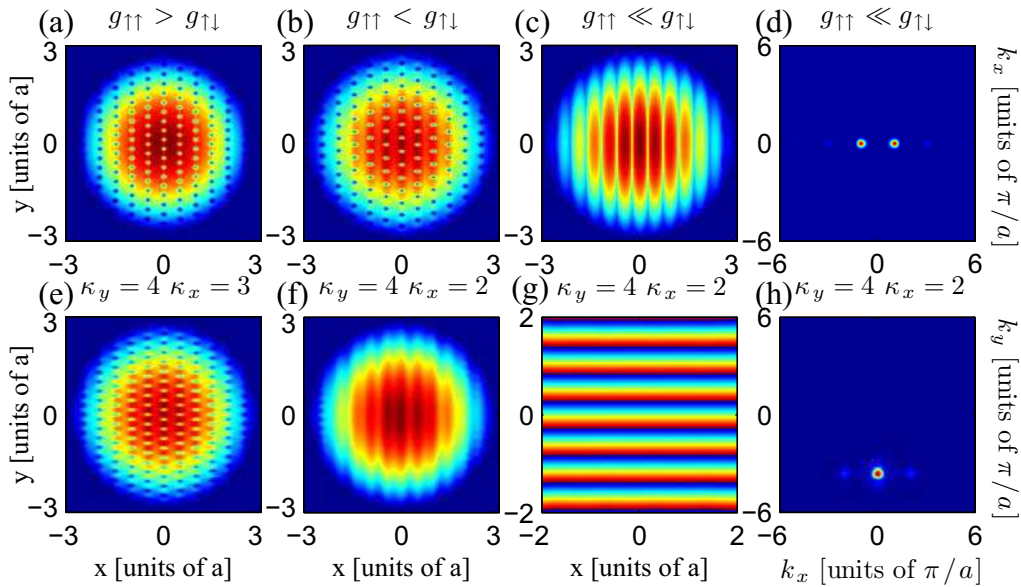}
\caption{(Color online) Effects of asymmetric interaction and
anisotropic spin-orbit coupling. (a-c) Density distributions of Bose
condensates with Rashba SO coupling and asymmetric interaction.
While the intra-component interaction is fixed with
$\tilde{g}_{\uparrow\uparrow}=\tilde{g}_{\downarrow\downarrow}=6000$,
the inter-component interactions are varied to be (a)
$\tilde{g}_{\uparrow\downarrow}=3000$, (b)
$\tilde{g}_{\uparrow\downarrow}=7000$, and (c)
$\tilde{g}_{\uparrow\downarrow}=18000$. The Rashba SO-coupling
strength is taken as  $\kappa=4\pi\hbar/ma$ in these plots. Notice
that a sufficiently strong inter-component interaction can drive the
system from the supersolid phase to the superfluid phase. (e, f)
Density distributions of Bose condensates in the presence of
anisotropic SO interaction with $\kappa_{x} \neq \kappa_y$.
Parameters used in these two plots are (e)
$\kappa_{x}=3\pi\hbar/ma$, $\kappa_{y}= 4\pi\hbar/ma$ and (f)
$\kappa_{x}= 2\pi\hbar/ma$, $\kappa_{y}= 4\pi\hbar/ma$. Here, the
interatomic interaction are considered to be SU(2) symmetric with
$\tilde{g} = 6000$. By increasing the SO-coupling anisotropy beyond
a certain value, the system will undergo a phase transition and
become a superfluid which can be regarded as a plane-wave state
characterized by phase modulation shown in (g). We emphasize that
the superfluid phases driven by asymmetric interaction (c) and by
anisotropic SO coupling (f) acquire different properties, as can be
easily distinguished from the momentum distribution depicted in (d)
and (h), respectively. In this figure, the periodic potential depth
is taken as $V_{0}=20\pi^2\hbar^2/ma^2$.} \label{fig6}
\end{figure*}

The relative phase plays an important role in determining the type
of the domain walls, which separate the spin-up and spin-down
domains \cite{Malomed2}. From the definition of the spin density
vector $\mathbf{S}$, one finds that $S_{z}$ is uniquely determined
by the relative density, while the direction of the spin projection
on the $S_{x}$-$S_{y}$ plane is determined by the relative phase and
can be represented by an azimuthal angle
$\alpha=\arctan(S_{y}/S_{x})=\theta_{\downarrow}-\theta_{\uparrow}$.
In the absence of SO coupling, the two component condensates have an
arbitrary relative phase, such that the spin projection on the
$S_{x}$-$S_{y}$ plane within the domain wall can take arbitrary
directions. In the presence of SO coupling, the relative phase is
locked at $\pm\pi/2$, thus the spin rotational symmetry in the
$S_{x}$-$S_{y}$ plane is broken. As $S_{x}=2|\Psi _{\uparrow}||\Psi
_{\downarrow}|\cos(\theta _{\uparrow}-\theta _{\downarrow})/(|\Psi
_{\uparrow}|^2+|\Psi _{\downarrow}|^2)$, obviously we have $S_x=0$.
This implies that the spins on the domain wall are confined within
the $S_{y}$-$S_{z}$ plane and form a Bloch wall, crossing which the
spin vector rotates like a spiral~\cite{Chen}.

One important feature of a domain wall is its chirality, which
distinguishes the right-handed rotation from the left-handed
rotation as moving between domains. Domain wall chirality has been
recently investigated in ultrathin ferromagnetic
films~\cite{Chen,Ryu,Emori}. As a new controllable degree of
freedom, domain wall chirality opens up new opportunities for
spintronics device designs, and has potential application in
information processing and storage. In the present system, we find
that the chirality of the Bloch walls can be manipulated by changing
the sign of the Rashba SO coupling. According to Eq.~(\ref{SO
coupling term}), if one changes the sign of the Rashba SO-coupling
constant, the relative phase will jump between $\pm\pi/2$ [see
Fig.~\ref{fig5}(a)]. As $S_{y}=-2|\Psi _{\uparrow}||\Psi
_{\downarrow}|\sin(\theta _{\uparrow}-\theta _{\downarrow})/(|\Psi
_{\uparrow}|^2+|\Psi _{\downarrow}|^2)$, changing the sign of the
relative phase will change the sign of $S_{y}$, and hence the
chirality of the Bloch walls. Typical examples of the spin
configurations are given in Figs.~\ref{fig5}(b) and \ref{fig5}(c), where the
right-handed and left-handed chiral Bloch walls correspond to
positive and negative Rashba SO-coupling constants, respectively. In
a realistic experiment, the sign of the Rashba SO coupling can be
varied by tuning the phase of the rf field~\cite{Anderson} in the
proposal described in Appendix \ref{App:Creat}.

\section{Effects of asymmetric interaction and anisotropic spin-orbit coupling}

In the discussion above, we have focused on the case of SU(2)
symmetric interaction with
$g_{\uparrow\uparrow}=g_{\downarrow\downarrow}=g_{\uparrow\downarrow}$.
It is important to consider also the non-SU(2) symmetric
interaction with $g_{\uparrow\uparrow}=g_{\downarrow\downarrow}\neq
g_{\uparrow\downarrow}$. We find that if a supersolid phase can be
stabilized with a proper combination of SO-coupling strength
$\kappa$ and periodic potential depth $V_{0}$ with a SU(2)
symmetric interaction, an asymmetric interaction with
$g_{\uparrow\uparrow} > g_{\uparrow\downarrow}$ always favors the
supersolid phase, as shown in Fig.~\ref{fig6}(a). The supersolid
phase is also stable for
$g_{\uparrow\uparrow}<g_{\uparrow\downarrow}$ provided that the
difference in
 $g_{\uparrow\downarrow}$ and $g_{\uparrow\uparrow}$ is sufficiently small,
$ g_{\uparrow\downarrow}-g_{\uparrow\uparrow}\ll
g_{\uparrow\downarrow}$. Such a situation corresponds to
Fig.~\ref{fig6}(b). As one further increases the asymmetry such that
$g_{\uparrow\uparrow}\ll g_{\uparrow\downarrow}$, the supersolid
phase becomes unfavorable and is replaced by the superfluid phase
[Figs.~\ref{fig6}(c) and \ref{fig6}(d)].

Additionally we have also considered the anisotropy effects of the
SO coupling. By decreasing $\kappa_x$ we find that the supersolid
phase, if it exists in the Rashba case, remains stable for a certain
range of $\kappa_x < \kappa_y$, as shown in Fig.~\ref{fig6}(e). By
further increasing the anisotropy, the system undergoes a phase
transition and becomes a superfluid, as shown in Fig.~\ref{fig6}(f).
This superfluid phase can be regarded as a plane-wave state
characterized by a phase modulation in the $y$ direction, as shown
in Fig. \ref{fig6}(g). The momentum distribution for this case is
illustrated in Fig. \ref{fig6}(h). In particular, when
$\kappa_{x}=0$, the SO coupling becomes unidirectional and reduces
to that of the National Institute of Standards and Technology scheme \cite{Spielman,Shuai-Chen,Shuai-Chen2},
and the supersolid phase with nontrivial topological spin texture is
no longer formed.

\section{Discussion}
The system considered can be realized experimentally in
$^{\text{87}}$Rb condensates using two magnetic states $\left\vert
F=1,m_{F}=1\right\rangle$\ and $ \left\vert
F=1,m_{F}=-1\right\rangle$ of the $F=1$ ground-state manifold. The
Rashba SO coupling and spin-dependent periodic potential can be
implemented by a combination of magnetic pulses \cite{Anderson,Xu},
and a pair of cross-linear polarized counterpropagating laser
beams~\cite{Deutsch,Mandel}. The detail proposal is given in
Appendix \ref{App:Creat}. Considering a typical experimental
situation in which a total of $N = 1.7 \times 10^5$ atoms with the
$s$-wave scattering length $a_{\alpha\beta}\approx100a_{B}$ ($a_{B}$
is the Bohr radius) are confined in a harmonic trap with the
frequencies $\omega _{\bot }\approx2\pi \times 40$ Hz and
$\omega_{z}\approx2\pi \times 200$~Hz, we obtain the effective
interaction parameter $\tilde{g}\approx 6000$. By using a
CO$_{\text{2}}$ laser operated at a wavelength of 10.6~$\mu$m, one
can produce a lattice constant $a$ coinciding with
$\pi\sqrt{\hbar/m\omega_{\perp}}$. These are consistent with the
parameters used in our calculations.

The supersolid phase can be identified either by a direct
observation of the lattice structure via {\it in-situ}
measurements~\cite{Chin,Greiner} or by momentum distribution
measurements using the time-of-fight imaging technique
~\cite{Sengstock}. The topological spin configurations of the
meron-pair textures, as well as the chiral domain walls, can be
imaged nondestructively with a high spatial resolution by the
magnetization-sensitive phase-contrast imaging technique
\cite{Sadler}. The domain wall chirality can also be determined by
extracting the relative phase from the dual state imaging technique
\cite{Anderson2}.

To summarize, we have studied the spin-orbit-coupled Bose-Einstein
condensates in a spin-dependent periodic potential. We have
demonstrated that the interplay between the spin-orbit coupling and
the spin-dependent periodic potential leads to the emergence of a
supersolid phase, which features a concomitant magnetic ordering
with topologically nontrivial spin textures. We have explored the
phase diagram of the system upon changing the spin-orbit-coupling
strength and the periodic potential depth, and investigated the
effects of asymmetric interatomic interaction and anisotropic
spin-orbit coupling. Proposals to realize and observe the supersolid
phase within realistic experimental situations have also been
discussed.

\section*{ACKNOWLEDGMENTS}
This work was supported by the NKBRSFC under Grants
No. 2011CB921502 and No. 2012CB821305; NSFC under
Grants No. 61227902, No. 61378017, No. 11274009, and
No. 11375030; SPRPCAS under Grant No. XDB01020300;
and the European Social Fund under the Global Grant measure.

\appendix
\renewcommand \appendixname{APPENDIX}

\section{CREATING SPIN-ORBIT-COUPLED BOSE GASES IN A SPIN-DEPENDENT PERIODIC POTENTIAL}\label{App:Creat}
\begin{figure*}[tbp]
\includegraphics[width=13cm]{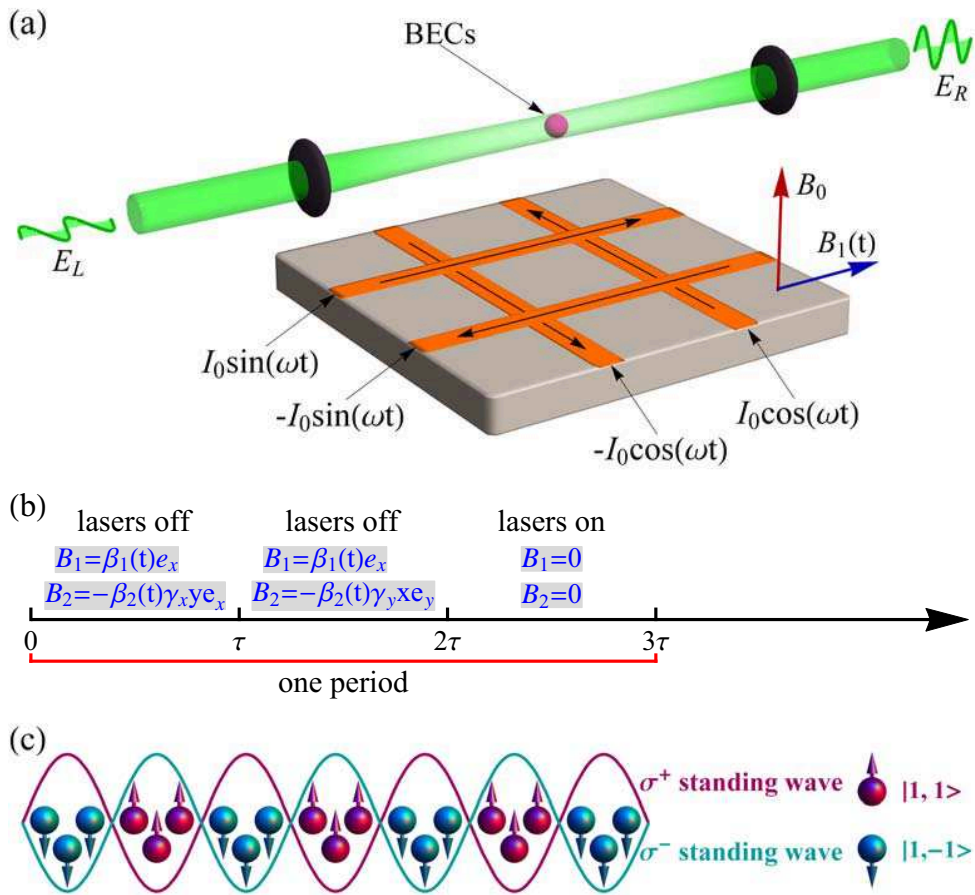}
\caption{(Color online) Experimental setup for creating spin-orbit-coupled Bose gases in a spin-dependent periodic potential. (a) The
cloud of atoms is situated tens of micrometers above the surface of
an atom chip. Two pairs of parallel microwires with amplitude
modulated rf current are embedded in the chip, and produce periodic
pulsed magnetic field gradients along perpendicular directions.
Another pulsed uniform magnetic field oriented along the $x$ axis is
added, with the sum frequency of the two magnetic fields equal to
the magnetic splitting induced by a strong bias field
$B_{0}\textbf{e}_{z}$ between the sublevels $m_F=\pm 1$. This
provides a two-photon coupling between the hyperfine states, and
induces an effective 2D SO coupling in the first-order approximation
to the pulse duration $\tau$. Two counterpropagating linearly
polarized laser beams with the same frequency but perpendicular
polarization vectors create a spin-dependent periodic potential. (b)
The pulse sequence used to implement SO coupling and spin-dependent
periodic potential. The parameters $\gamma_{x}$ and $\gamma_{y}$
characterize the strength of the magnetic field gradient, and
$\beta_{1}(t)$ and $\beta_{2}(t)$ define the temporal shape of the
magnetic fields. (c) Two polarized standing wave laser fields
$\sigma^{+}$ (purple) and $\sigma^{-}$ (cyan) are produced by the
counterpropagating linearly polarized lasers in (a). Due to the
polarization-dependent a.c. Stark shift, the internal state
$\left\vert 1,1\right\rangle$ is affected by the standing wave laser
field $\sigma^{+}$, while the internal state $\left\vert
1,-1\right\rangle$ experiences the standing wave laser field
$\sigma^{-}$.} \label{figS1}
\end{figure*}

We consider a two-component Bose gas of ultacold alkali atoms, such
as $^{\text{87}}$Rb, with two internal states chosen to be the
hyperfine states $\left\vert F=1,m_{F}=1\right\rangle$ and $
\left\vert F=1,m_{F}=-1\right\rangle$ of the $F=1$ ground-state
manifold. The protocol for implementing the spin-orbit (SO) coupling
and the spin-dependent periodic potential is illustrated in
Fig.~\ref{figS1}. It relies on the ability to switch between
magnetic pulses and laser pulses [see Fig.~\ref{figS1}(b)].

The first two stages of the scheme represent a modified version of a
recent proposal \cite{Anderson} to produce SO coupling by means of
magnetic pulses. Originally it was proposed to create the SO
coupling using a strong time-independent bias magnetic field along
the quantization axis $z$ and infrared~(IR) magnetic field in the
$x$-$y$ plane with a frequency $\omega$ in resonance with splitting
between the magnetic sublevels induced by the bias
field~\cite{Anderson}. Yet now we are dealing with the hyperfine
states $\left\vert F=1,m_{F}=1\right\rangle$ and $ \left\vert
F=1,m_{F}=-1\right\rangle$, which cannot be directly coupled by such
magnetic pulses. To bypass the problem, we propose to use
simultaneously two IR magnetic fields in the $x$-$y$ plane with
different frequencies $\omega_1\ne\omega_2$, where frequency sum
$\omega_1+\omega_2$ is equal to the magnetic splitting between the two
sublevels. This provides a two photon coupling between the hyperfine
states $\left\vert F=1,m_{F}=1\right\rangle$ and $ \left\vert
F=1,m_{F}=-1\right\rangle$. The corresponding second-order coupling
Hamiltonian can be made proportional to $\sigma_x$ or $\sigma_y$
depending on the phases of the IR fields, like in Ref.
\cite{Anderson}, where $\sigma_x$ and $\sigma_y$ are the quasi-spin
operators for the selected pair of states.

The magnetic field $\textbf{B}_{1}$ with frequency $\omega_1$ is
taken to be uniform and oriented along the $x$ axis. Another
magnetic field $\textbf{B}_{2}$ with frequency $\omega_2$ is
produced by a pair of wires along the $y$ or $x$ axis for the first
($0\leq t<\tau$) and the second ($\tau\leq t<2\tau$) stages,
respectively~\cite{Anderson}. By going to the rotating frame to
eliminate the bias field along the $z$ direction, choosing the
proper phases of the IR magnetic fields, and making the
rotating-wave-approximation to neglect the fast oscillating terms,
the second order coupling induced by the IR fields can yield the SO-coupling terms $-i\hbar\kappa_{x}\sigma_{x}\partial_{x}$ and
$-i\hbar\kappa_{y}\sigma_{y}\partial_{y}$ for the first and second
stages respectively. The SO-coupling parameters $\kappa_{x}$ and
$\kappa_{y}$ depend on the strength of the magnetic pulses and the
detuning from the single photon resonance, and also require some
quadratic Zeeman shift in order to be non-zero \cite{footnote1}. The
first two stages provide a 2D SO
coupling~\cite{Anderson,Xu}
$\mathcal{V}_{\text{so}}=-i\hbar(\kappa_{x}\sigma_{x}\partial_{x}+\kappa_{y}\sigma_{y}\partial_{y})$
in the first-order approximation, which is valid for a sufficiently
short duration $\tau$. In particular, for $\kappa_{x}=\kappa_{y}$,
one arrives at the isotropic Rashba-type SO coupling.

In the third stage, $2\tau\leq t<3\tau$, the magnetic field is
turned off, and two counterpropagating laser beams are applied with
the same frequency but perpendicular linear polarization vectors
[see Fig.~\ref{figS1}(a)]. In this case, a standing wave light field
is formed. It can be decomposed into a superposition of $\sigma^{+}$
and $\sigma^{-}$ polarized standing waves, giving rise to periodic
potentials $V_{+}=V_{0}\sin^{2}(\pi x/a)$ and
$V_{-}=V_{0}\cos^{2}(\pi x/a)$ \cite{Mandel}. Due to the
polarization-dependent a.c. Stark shift, atoms in different
hyperfine states will feel significantly different
potentials~\cite{Grimm}. For the $F=1$ ground-state manifold chosen
above, the internal state $\left\vert F=1,m_{F}=1\right\rangle$
experiences the $V_{+}$ potential and the internal state $\left\vert
F=1,m_{F}=-1\right\rangle$ is affected by the $V_{-}$ potential.
This leads to the formation of the spin-dependent periodic potential
\cite{McKay}, as shown in Fig.~\ref{figS1}(c).

\section{CALCULATING THE MANY-BODY GROUND STATES}\label{App:Calculat}
We investigate the many-body effects based on the Gross-Pitaevskii
mean-field theory. It is well expected that a mean-field approach is
valid provided that the system is far away from a quantum critical
point such that the quantum fluctuation effects are not significant.
In the case of one-dimensional optical lattice where each lattice
site is essentially a one-dimensional tube containing a large number
of particles, the Wannier function can be drastically altered from
the single-particle form by the interaction effect. As a
consequence, the critical value of lattice depth is dependent on the
atom number on each lattice site~\cite{li-03}. For the parameters
considered in this manuscript, the particle number in each tube is
as high as several thousand, which ensures that the critical value
of lattice depth is above $90$ recoil energy. Thus, we expect the
mean-field approach to give a satisfactory description of the system
for $V_0 \lesssim 90$ recoils.

The validity of the Gross-Pitaevskii mean-field approximation used
above can be checked by evaluating the quantum depletion caused by
quantum fluctuations~\cite{Pethick}. According to the Bogoliubov
theory, the fluctuation part
$\delta\hat{\Psi}_{\alpha}(\mathbf{r},t)$ around the condensate can
be subjected to a canonical transformation resulting in the
expansion $\delta\hat{\Psi}_{\alpha}(\mathbf{r},t)=\sum_{j}\big[
u_{\alpha j}(\mathbf{r}) e^{-i\omega_{j}t}\hat{%
\gamma}_{j}+v_{\alpha j}^{\ast }(\mathbf{r}) e^{i\omega _{j}t}
\hat{\gamma}_{j}^{\dag }\big]$, where $\hat{\gamma}_{j}$ and
$\hat{\gamma}_{j}^{\dag}$ are the quasiparticle creation and
annihilation operators associated with the $j$th collective mode.
The mode functions $u_{\alpha i}(\mathbf{r})$, $v_{\alpha
i}(\mathbf{r})$ and collective frequencies $\omega_{j}$ are
determined by the Bogoliubov-de Gennes (BdG) equations
\begin{widetext}
\begin{equation} \label{Bogoliubov-de Gennes equations}
\left[
\begin{array}{cccc}
H_{s_{\uparrow }}+g_{\uparrow \uparrow }\left\vert \Psi _{\uparrow
}\right\vert ^{2} & V_{\text{so}}+g_{\uparrow \downarrow }\Psi
_{\uparrow }\Psi _{\downarrow }^{\ast } & g_{\uparrow \uparrow }\Psi
_{\uparrow }^{2} &
g_{\uparrow \downarrow }\Psi _{\uparrow }\Psi _{\downarrow } \\
-V_{\text{so}}^{\ast }+g_{\downarrow \uparrow }\Psi _{\downarrow
}\Psi _{\uparrow }^{\ast } & H_{s_{\downarrow }}+g_{\downarrow
\downarrow }\left\vert \Psi _{\downarrow }\right\vert ^{2} &
g_{\downarrow \uparrow }\Psi _{\downarrow }\Psi _{\uparrow } &
g_{\downarrow \downarrow }\Psi
_{\downarrow }^{2} \\
g_{\uparrow \uparrow }\Psi _{\uparrow }^{\ast 2} & g_{\uparrow
\downarrow }\Psi _{\uparrow }^{\ast }\Psi _{\downarrow }^{\ast } &
H_{s_{\uparrow
}}+g_{\uparrow \uparrow }\left\vert \Psi _{\uparrow }\right\vert ^{2} & V_{%
\text{so}}^{\ast }+g_{\uparrow \downarrow }\Psi _{\uparrow }^{\ast
}\Psi
_{\downarrow } \\
g_{\downarrow \uparrow }\Psi _{\downarrow }^{\ast }\Psi _{\uparrow
}^{\ast }
& g_{\downarrow \downarrow }\Psi _{\downarrow }^{\ast 2} & -V_{\text{so}%
}+g_{\downarrow \uparrow }\Psi _{\downarrow }^{\ast }\Psi _{\uparrow
} & H_{s_{\downarrow }}+g_{\downarrow \downarrow }\left\vert \Psi
_{\downarrow
}\right\vert ^{2}%
\end{array}%
\right] \left[
\begin{array}{c}
u_{\uparrow i}\left( \mathbf{r}\right)  \\
u_{\downarrow i}\left( \mathbf{r}\right)  \\
v_{\uparrow i}\left( \mathbf{r}\right)  \\
v_{\downarrow i}\left( \mathbf{r}\right)
\end{array}%
\right] =\hbar \omega _{i}\left[
\begin{array}{c}
u_{\uparrow i}\left( \mathbf{r}\right)  \\
u_{\downarrow i}\left( \mathbf{r}\right)  \\
-v_{\uparrow i}\left( \mathbf{r}\right)  \\
-v_{\downarrow i}\left( \mathbf{r}\right)
\end{array}%
\right],
\end{equation}
\end{widetext}
under the normalization $\int\big( \left\vert u_{\uparrow
i}\right\vert ^{2}+\left\vert u_{\downarrow i}\right\vert
^{2}-\left\vert v_{\uparrow i}\right\vert ^{2}-\left\vert
v_{\downarrow i}\right\vert ^{2}\big) d\mathbf{r}=1$. Here,
$H_{s_{\uparrow }}=H_{\text{osc}}+V_{\uparrow }+g_{\uparrow \uparrow
}\left\vert \Psi _{\uparrow }\right\vert ^{2}+g_{\uparrow \downarrow
}\left\vert \Psi _{\downarrow }\right\vert ^{2}-\mu$,
$H_{s_{\downarrow }}=H_{\text{osc}}+V_{\downarrow }+g_{\downarrow
\downarrow }\left\vert \Psi _{\downarrow }\right\vert
^{2}+g_{\downarrow \uparrow }\left\vert \Psi _{\uparrow }\right\vert
^{2}-\mu $ with $H_{\text{osc}}=-\frac{\hbar ^{2}}{2m}\mathbf{\nabla
}^{2}+V_{\text{H}}$ and $\mu$ the chemical potential, and
$V_{\text{so}}=-\hbar(i \kappa_{x} \partial _{x}+\kappa_{y} \partial
_{y})$. At zero temperature, the number of the non-condensate
particles can be calculated by $\delta N=\int\sum_{j}\big(
\left\vert v_{\uparrow i}\right\vert ^{2}+\left\vert v_{\downarrow
i}\right\vert ^{2}\big) d\mathbf{r}$, where $j$ is restricted by the
nonnegative mode frequencies $\omega_{j}>0$.

\begin{figure}[tb]
\includegraphics[width=8.5cm]{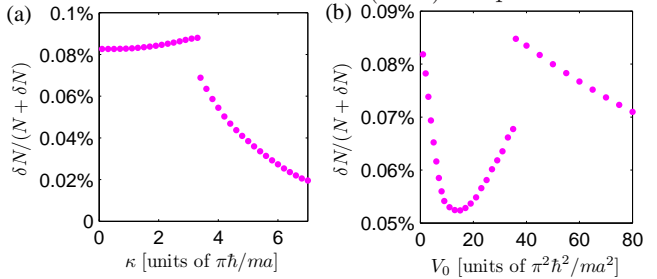}
\caption{(Color online) Quantum depletion as a function of the
spin-orbit-coupling strength (a) and the periodic potential depth
(b). The periodic potential depth is fixed at
$V_{0}=20\pi^2\hbar^2/ma^2$ in (a), and the spin-orbit-coupling
strength is fixed at $\kappa=4\pi\hbar/ma$ in (b). The dimensionless
interaction parameter is taken as $\tilde{g}=6000$. The quantum
depletion is always less than $0.1\%$, thereby confirming the
validity of the mean-field approach.} \label{figS2}
\end{figure}

By numerically solving the BdG equations~(\ref{Bogoliubov-de Gennes
equations}) in two dimension, we find that the quantum depletion is
small not only in the superfluid phase but also in the supersolid
phase, thus the quantum fluctuations can be neglected. In
Figs.~\ref{figS2}(a) and \ref{figS2}(b), we present the quantum
depletion $\delta N/(N+\delta N)$ as a function of the SO-coupling
strength $\kappa$ and the periodic potential depth $V_{0}$,
respectively. One can see that, the quantum depletion is always less
than $0.1\%$, thereby confirming the validity of the
Gross-Pitaevskii approach.

By numerically minimizing the energy functional, we can obtain the
many-body ground-state wave functions. A valid and widely used
method for the minimization is the imaginary time algorithm
\cite{Dalfovo,Chiofalo}. In solving the imaginary time evolution
equations, we develop a backward-forward Euler
Fourier-pseudospectral (BFFP) discretization. For the time
discretization, we use the backward or forward Euler scheme for
linear or nonlinear terms in time derivatives. For the spatial
discretization, we take fast Fourier transform (FFT) in spatial
derivatives. A similar discretization scheme, named backward-forward
Euler sine-pseudospectral (BFSP) discretization, has been proposed
and demonstrated for Bose systems without SO coupling~\cite{Bao}.


\begin{thebibliography}{99}

\bibitem{Andreev} A. F. Andreev and I. M. Lifshitz, Sov. Phys. JETP \textbf{29}, 1107 (1969).

\bibitem{Chester} G. V. Chester, Phys. Rev. A \textbf{2}, 256 (1970).

\bibitem{Leggett} A. J. Leggett, Phys. Rev. Lett. \textbf{25}, 1543 (1970).

\bibitem{Meisel} M. W. Meisel, Physica B \textbf{178}, 121 (1992).

\bibitem{Balibar} S. Balibar and F. Caupin, J. Phys.: Condens. Matter \textbf{20}, 173201 (2008).

\bibitem{Boninsegni} M. Boninsegni and N. V. Prokof'ev, Rev. Mod. Phys. \textbf{84}, 759 (2012).

\bibitem{Balibar2} S. Balibar, Nature (London) \textbf{464}, 176 (2010).

\bibitem{Kim} E. Kim and M. H. W. Chan, Nature (London) \textbf{427}, 225 (2004).

\bibitem{Ye} J. Ye, Phys. Rev. Lett. \textbf{97}, 125302 (2006).

\bibitem{Day} J. Day and J. Beamish, Nature (London) \textbf{450}, 853 (2007).

\bibitem{Hunt} B. Hunt, E. Pratt, V. Gadagkar, M. Yamashita, A. V. Balatsky, and J. C. Davis, Science \textbf{324}, 632 (2009).

\bibitem{Kim2} H. Choi, D. Takahashi, K. Kono, and E. Kim, Science \textbf{330}, 1512 (2010).

\bibitem{Chan} D. Y. Kim and M. H. W. Chan, Phys. Rev. Lett. \textbf{109}, 155301 (2012).

\bibitem{Henkel} N. Henkel, R. Nath, and T. Pohl, Phys. Rev. Lett. \textbf{104}, 195302 (2010).

\bibitem{Cinti} F. Cinti, P. Jain, M. Boninsegni, A. Micheli, P. Zoller, and G. Pupillo, Phys. Rev. Lett. \textbf{105}, 135301 (2010).

\bibitem{Henkel2} N. Henkel, F. Cinti, P. Jain, G. Pupillo, and T. Pohl, Phys. Rev. Lett. \textbf{108}, 265301 (2012).

\bibitem{Troyer} S. Wessel and M. Troyer, Phys. Rev. Lett. \textbf{95}, 127205 (2005).

\bibitem{Danshita} I. Danshita and C. A. R. S\'{a} de Melo, Phys. Rev. Lett. \textbf{103}, 225301 (2009).

\bibitem{Tieleman} O. Tieleman, A. Lazarides, and C. Morais Smith, Phys. Rev. A \textbf{83}, 013627 (2011).

\bibitem{Boninsegni2} M. Boninsegni, J. Low. Temp. Phys. \textbf{168}, 137 (2012).

\bibitem{Lewenstein} M. Lewenstein, A. Sanpera, V. Ahufinger, B. Damski, A. Sen(De), and U. Sen, Adv. Phys. \textbf{56}, 243 (2007).

\bibitem{Bloch} I. Bloch, J. Dalibard, and W. Zwerger, Rev. Mod. Phys. \textbf{80}, 885 (2008).

\bibitem{Spielman} Y. J. Lin, K. Jim\'{e}nez-Garc\'{\i}a, and I. B. Spielman, Nature (London) \textbf{471}, 83 (2011).

\bibitem{Shuai-Chen} J. Y. Zhang, S. C. Ji, Z. Chen, L. Zhang, Z. D. Du, B. Yan, G. S. Pan, B. Zhao, Y. J. Deng, H. Zhai, S. Chen, and J. W. Pan, Phys. Rev. Lett. \textbf{109}, 115301 (2012).

\bibitem{Shuai-Chen2} S. C. Ji, J. Y. Zhang, L. Zhang, Z. D. Du, W. Zheng, Y. J. Deng, H. Zhai, S. Chen, and J. W. Pan, Nat. Phys. \textbf{10}, 314 (2014).

\bibitem{Jing-Zhang} P. Wang, Z. Q. Yu, Z. Fu, J. Miao, L. Huang, S. Chai, H. Zhai, and J. Zhang, Phys. Rev. Lett. \textbf{109}, 095301 (2012).

\bibitem{Zwierlein} L. W. Cheuk, A. T. Sommer, Z. Hadzibabic, T. Yefsah, W. S. Bakr, and M. W. Zwierlein, Phys. Rev. Lett. \textbf{109}, 095302 (2012).

\bibitem{Engels} C. Qu, C. Hamner, M. Gong, C. Zhang, and P. Engels, Phys. Rev. A \textbf{88}, 021604(R) (2013).

\bibitem{Jing-Zhang2} Z. Fu, L. Huang, Z. Meng, P. Wang, L. Zhang, S. Zhang, H. Zhai, P. Zhang, and J. Zhang, Nat. Phys. \textbf{10}, 110 (2014).

\bibitem{Dalibard} J. Dalibard, F. Gerbier, G. Juzeli\={u}nas, and P. \"{O}hberg, Rev. Mod. Phys. \textbf{83}, 1523 (2011).

\bibitem{Galitski} V. Galitski and I. B. Spielman, Nature (London) \textbf{494}, 49 (2013).

\bibitem{Wu} X. Zhou, Y. Li, Z. Cai, and C. Wu, J. Phys. B \textbf{46}, 134001 (2013).

\bibitem{Niu} A. M. Dudarev, R. B. Diener, I. Carusotto, and Q. Niu, Phys. Rev. Lett. \textbf{92}, 153005 (2004).

\bibitem{Galitski2} T. D. Stanescu, B. Anderson, and V. Galitski, Phys. Rev. A \textbf{78}, 023616 (2008).

\bibitem{Wu2} C. Wu, I. Mondragon-Shem, and X. F. Zhou, Chin. Phys. Lett. \textbf{28}, 097102 (2011).

\bibitem{Wuming-Liu} R. Liao, Y. Yi-Xiang, and W. M. Liu, Phys. Rev. Lett. \textbf{108}, 080406 (2012).

\bibitem{Trivedi} W. S. Cole, S. Zhang, A. Paramekanti, and N. Trivedi, Phys. Rev. Lett. \textbf{109}, 085302 (2012).

\bibitem{Wei-Zhang} W. Zhang and W. Yi, Nat. Commun. \textbf{4}, 2711 (2013).

\bibitem{Hui-Zhai} C. Wang, C. Gao, C. M. Jian, and H. Zhai, Phys. Rev. Lett. \textbf{105}, 160403 (2010).

\bibitem{Tin-Lun-Ho} T. L. Ho and S. Zhang, Phys. Rev. Lett. \textbf{107}, 150403 (2011).

\bibitem{Malomed} H. Sakaguchi, B. Li, and B. A. Malomed, Phys. Rev. E \textbf{89}, 032920 (2014).

\bibitem{You} Z. F. Xu, R. L\"{u}, and L. You, Phys. Rev. A \textbf{83}, 053602 (2011).

\bibitem{Hu} H. Hu, B. Ramachandhran, H. Pu, and X. J. Liu, Phys. Rev. Lett. \textbf{108}, 010402 (2012).

\bibitem{Santos} S. Sinha, R. Nath, and L. Santos, Phys. Rev. Lett. \textbf{107}, 270401 (2011).

\bibitem{Gou} S. W. Su, I. K. Liu, Y. C. Tsai, W. M. Liu, and S. C. Gou, Phys. Rev. A \textbf{86}, 023601 (2012).

\bibitem{Su-Yi} Y. Deng, J. Cheng, H. Jing, C. P. Sun, and S. Yi, Phys. Rev. Lett. \textbf{108}, 125301 (2012).

\bibitem{Gopalakrishnan} S. Gopalakrishnan, I. Martin, and E. A. Demler, Phys. Rev. Lett. \textbf{111}, 185304 (2013).

\bibitem{Ye2} Y. Chen, J. Ye, and G. Tian, J. Low. Temp. Phys. \textbf{169}, 149 (2012).

\bibitem{Pitaevskii} Y. Li, G. I. Martone, L. P. Pitaevskii, and S. Stringari, Phys. Rev. Lett. \textbf{110}, 235302 (2013).

\bibitem{Deutsch} I. H. Deutsch and P. S. Jessen, Phys. Rev. A \textbf{57}, 1972 (1998).

\bibitem{Mandel} O. Mandel, M. Greiner, A. Widera, T. Rom, T. W. H\"{a}nsch, and I. Bloch, Phys. Rev. Lett. \textbf{91}, 010407 (2003).

\bibitem{Goldman} N. Goldman, G. Juzeli\={u}nas, P. \"{O}hberg, and I. B. Spielman, Rep. Prog. Phys. \textbf{77}, 126401 (2014).

\bibitem{Anderson} B. M. Anderson, I. B. Spielman, and G. Juzeli\={u}nas, Phys. Rev. Lett. \textbf{111}, 125301 (2013).

\bibitem{Xu} Z. F. Xu, L. You, and M. Ueda, Phys. Rev. A \textbf{87}, 063634 (2013).

\bibitem{Pethick} C. J. Pethick and H. Smith, {\it Bose-Einstein Condensation in Dilute Gases} (Cambridge University Press, Cambridge, 2002).

\bibitem{Volovik} G. E. Volovik, {\it The Universe in a Helium Droplet} (Oxford University Press, New York, 2003).

\bibitem{Kasamatsu} K. Kasamatsu, M. Tsubota, and M. Ueda, Phys. Rev. Lett. \textbf{93}, 250406 (2004).

\bibitem{Schweikhard} V. Schweikhard, I. Coddington, P. Engels, S. Tung, and E. A. Cornell, Phys. Rev. Lett. \textbf{93}, 210403 (2004).

\bibitem{Gou2} S. W. Su, C. H. Hsueh, I. K. Liu, T. L. Horng, Y. C. Tsai, S. C. Gou, and W. M. Liu, Phys. Rev. A \textbf{84}, 023601 (2011).

\bibitem{Malomed2} B. A. Malomed, H. E. Nistazakis, D. J. Frantzeskakis, and P. G. Kevrekidis, Phys. Rev. A \textbf{70}, 043616 (2004).

\bibitem{Chen} G. Chen, T. Ma, A. T. N'Diaye, H. Kwon, C. Won, Y. Wu, and A. K. Schmid, Nat. Commun. \textbf{4}, 2671 (2013).

\bibitem{Ryu} K. S. Ryu,  L. Thomas, S. H. Yang, and S. Parkin, Nat. Nanotechnol. \textbf{8}, 527 (2013).

\bibitem{Emori} S. Emori, U. Bauer, S. M. Ahn, E. Martinez, and G. S. D. Beach, Nat. Mater. \textbf{12}, 611 (2013).

\bibitem{Chin} N. Gemelke, X. Zhang, C. L. Hung, and C. Chin, Nature (London) \textbf{460}, 995 (2009).

\bibitem{Greiner} W. S. Bakr, A. Peng, M. E. Tai, R. Ma, J. Simon, J. I. Gillen, S. F\"{o}lling, L. Pollet, and M. Greiner, Science \textbf{329}, 547 (2010).

\bibitem{Sengstock} P. T. Ernst, S. G\"{o}tze, J. S. Krauser, K. Pyka, D. S. L\"{u}hmann, D. Pfannkuche, and K. Sengstock, Nat. Phys. \textbf{6}, 56 (2010).

\bibitem{Sadler} L. E. Sadler, J. M. Higbie, S. R. Leslie, M. Vengalattore, and D. M. Stamper-Kurn, Nature (London) \textbf{443}, 312 (2006).

\bibitem{Anderson2} R. P. Anderson, C. Ticknor, A. I. Sidorov, and B. V. Hall, Phys. Rev. A \textbf{80}, 023603 (2009).

\bibitem{footnote1} In the absence of the quadratic Zeeman shift, the two-photon
transitions induced first by magnetic field $\textbf{B}_{1}$
followed by $\textbf{B}_{2}$ interfere destructively with the
transitions induced by the magnetic fields  in the opposite order:
$\textbf{B}_{2}$ followed by $\textbf{B}_{1}$.

\bibitem{Grimm} R. Grimm, M. Weidem\"{u}ller, and Y. B. Ovchinnikov,  Adv. At., Mol., Opt. Phys. \textbf{42}, 95 (2000).

\bibitem{McKay} D. McKay and B. DeMarco, New J. Phys. \textbf{12}, 055013 (2010).

\bibitem{li-03} J. Li, Y. Yu, A. M. Dudarev, and Q. Niu, New J. Phys. \textbf{8}, 154 (2006).

\bibitem{Dalfovo} F. Dalfovo and S. Stringari, Phys. Rev. A \textbf{53}, 2477 (1996).

\bibitem{Chiofalo} M. L. Chiofalo, S. Succi, and M. P. Tosi, Phys. Rev. E \textbf{62}, 7438 (2000).

\bibitem{Bao} W. Bao, I. L. Chern, and F. Y. Lim, J. Comput. Phys. \textbf{219}, 836 (2006).

\end{thebibliography}
\end{document}